\documentclass[a4paper,12pt]{article}

\usepackage[a4paper, total={6.5in, 9in}]{geometry}
\usepackage[american]{babel}
\usepackage{latexsym}
\usepackage{float}
\usepackage{amsmath}
\usepackage{amssymb}
\usepackage{caption}
\usepackage{graphicx}
\usepackage[dvipsnames]{xcolor}
\usepackage{bm}
\usepackage{booktabs}
\usepackage{fancyvrb}
\usepackage{natbib}
\usepackage{setspace}
\usepackage{hyperref}
\usepackage{enumerate}

\newcommand{\norm}[1]{\left\lVert#1\right\rVert}
\newenvironment{keywords}{\noindent\textbf{keywords:}}{}
\DeclareMathOperator*{\argmin}{arg\,min}
\DeclareMathOperator*{\argmax}{arg\,max}

\title{\textbf{Fast Partial Quantile Regression}}

\author{Álvaro Méndez Civieta\thanks{Department of Statistics, Universidad Carlos III de Madrid.}\,{\ }\thanks{uc3m-Santander Big Data Institute.} \and M. Carmen Aguilera-Morillo\thanks{Department of Applied Statistics and Operational Research, and Quality, Universitat Politècnica de València}\,{\ }\footnotemark[2] \and Rosa E. Lillo\footnotemark[1]\,{\ }\footnotemark[2]}

\date{}

\begin{document}
\maketitle

\begin{abstract}

Partial least squares (PLS) is a dimensionality reduction technique used as an alternative to ordinary least squares (OLS) in situations where the data is colinear or high dimensional. Both PLS and OLS provide mean based estimates, which are extremely sensitive to the presence of outliers or heavy tailed distributions. In contrast, quantile regression is an alternative to OLS that computes robust quantile based estimates. In this work, the multivariate PLS is extended to the quantile regression framework, obtaining a theoretical formulation of the problem and a robust dimensionality reduction technique that we call fast partial quantile regression (fPQR), that provides quantile based estimates. An efficient implementation of fPQR is also derived, and its performance is studied through simulation experiments and the chemometrics well known biscuit dough dataset, a real high dimensional example.

\end{abstract}


\section{Introduction}\label{sec:intro}

Partial least squares (PLS) \citep{Wold1973}, \citep{Wold2001} is a dimensionality reduction technique commonly applied to two data blocks (predictors and responses) that works by projecting the available data into a latent structure. The key idea behind PLS is that it can summarize the predictors into a small set of uncorrelated latent variables that have maximal covariance with the responses. PLS has proven to be a versatile alternative to ordinary least squares (OLS), obtaining parsimonious models even when dealing with ill-posed multicollinear problems, commonly found in different areas of scientific research such as chemometrics, social science or medicine. See for example \citep{Nguyen2002}, where it is used in a tumor classification problem. In recent years PLS has also received attention when dealing with the increasingly common problem of high dimensional data, in which the number of observations is small and the number of variables is very large. In this regard, \cite{Boulesteix2006} successfully applied PLS to a genomic dataset. Partial least squares is based on the cross-covariance matrix between predictors and response, and on least squares models. Least squares models are known to behave nicely when the errors are normally distributed, but there is no guarantee that the normality will be satisfied in many experimental data problems, where heavy tailed distributions, and even outliers are expected to be found. This makes PLS extremely sensitive to the presence of outliers or non normal data. The solution to this problem has traditionally been centered in robustifying the least squares estimator in which PLS is based, see for example \citep{Serneels2005} where they make use of a robust M-regression estimator, or \citep{Acitas2020}, where a partial robust adaptive modified maximum likelihood estimator is proposed, among others.

Quantile regression \citep{Koenker1978} is an important statistical methodology that allows to describe the conditional quantiles of a response given a set of covariates. Fitting the data at a set of quantiles provides a more comprehensive picture of the response distribution than does the mean, and as opposed to least squares, quantile regression is resistant to outliers, and can deal with heavy tailed distributions and heteroscedasticity, the situation when variances depend on some covariates. Specifically, when the center of the distribution is of interest, the least absolute deviation (LAD), also called median regression, a particular case of quantile regression, provides more robust estimators than least squares regression. In recent years many papers have been published extending quantile regression to the high dimensional framework by performing variable selection, see for example \cite{Wu2009} where an adaptive lasso for quantile regression is introduced, or \citep{Mendez-Civieta2020}, where an adaptive sparse group lasso for quantile regression is proposed. However, to the best of our knowledge there is very little work on quantile based dimension reduction techniques. A well known PLS implementation is given by the NIPALS algorithm \citep{Wold1973}. \cite{Dodge2009} extended the NIPALS algorithm for univariate response problems to the quantile regression framework. They proposed a quantile covariance metric based on the quantile regression slope and used this metric to modify the univariate NIPALS, a modification that they called partial quantile regression (PQR). The work from \cite{Dodge2009} lays the foundation for an extension of PLS to the quantile regression framework, however we find some shortcomings in the development of the methodology and the algorithmic implementation that should be addressed. First, it has no background on what is the optimization problem that their PQR algorithm is solving. Second, it is centered in univariate response problems, providing no solution for multivariate response problems commonly found in fields such as chemometrics. Third, the computation time of their quantile coviariance, key in the algorithmic implementation, grows linearly with the number of variables, making solving high dimensional problems computationally expensive. The main contribution of our work is centered in addressing these problems. We define the optimization problem that the fPQR algorithm solves and study different quantile covariance alternatives \citep{Li2015b}, \citep{Choi2018}. We provide an efficient implementation of fPQR, greatly reducing the computation time when compared with that of \citep{Dodge2009} while achieving more accurate predictions. We also provide an implementation suitable for multivariate response settings. The result is a methodology that parallels the nice properties of PLS: it is a dimension reduction technique that obtains uncorrelated scores maximizing the quantile covariance between predictors and responses. But additionally, it is also a robust, quantile based methodology suitable for dealing with outliers, heteroscedastic or heavy tailed datasets. The median estimator of the fPQR algorithm is a robust alternative to PLS, while other quantile levels can provide additional information on the tails of the responses.

The rest of the paper is organized as follows. In Section \ref{sec:pls_intro} a brief introduction of the PLS algorithm for multivariate response is provided. Section \ref{sec:fPQR_definition} introduces the fPQR algorithm and studies different options for a quantile covariance metric. Section  \ref{sec:numerical_sim} tests the performance of the proposed fPQR algorithm in three synthetic dataset frameworks studying the quality of the estimated $\beta$ coefficients and the prediction error. In Section \ref{sec:real_data}, the proposed algorithm is used in a real high dimensional data example. Some computational aspects are briefly commented in Section \ref{sec:comput_aspect}, and the conclusions are provided in Section \ref{sec:conclusion}.


\section{The PLS model for multivariate response}\label{sec:pls_intro}

Let $X\in\mathbb{R}^{n\times m}$ and $Y\in\mathbb{R}^{n\times l}$ be two data matrices, samples drawn from some unknown population following the linear model,
\begin{equation}\label{eq:simple_lm}
    \bm{y}_i = \bm{x_i}B+\bm{\varepsilon}_i, \quad i=1,\ldots,n,
\end{equation}
where $\bm{y}_i\equiv(y_{i1},\ldots,y_{il})$ is the vector containing the response variables for the $i$-th observation, $\bm{x}_i\equiv(x_{i1},\ldots,x_{im})$ contains the predictive variables, $B\in\mathbb{R}^{m\times l}$ is the matrix containing the coefficients from the linear relations, and $\bm{\varepsilon_i}\equiv(\varepsilon_{i1},\ldots,\varepsilon_{il})$ is the error term. Without loss of generality, consider that both $X$ and $Y$ are mean centered. The PLS regression methodology works by assuming the existence of a latent structure,
\begin{equation}
X=TP^{t}+E; \quad Y=TQ^t+F,
\end{equation}
where $T\in\mathbb{R}^{n\times h}$ is the scores matrix formed by $h$ (usually being $h \ll m$) linear combinations of the original variables, $P\in\mathbb{R}^{m\times h}$ and $Q\in\mathbb{R}^{l\times h}$ are loadings matrices and $E\in\mathbb{R}^{n\times m}$ and $F\in\mathbb{R}^{n\times l}$ are random error matrices. The aim of PLS regression is precisely to regress the response matrix $Y$ onto the $h$ latent variables, stored in the scores matrix $T$, defining this way a low-dimensional regression model,
\begin{equation}\label{eq:lm_pls}
    \bm{y}_i = \bm{t_i}\Gamma+\bm{\varepsilon}_i^{*}, \quad i=1,\ldots,n,
\end{equation}
where $\Gamma$ is the matrix of regression coefficients. PLS is an iterative algorithm in which the scores in $T$ are obtained sequantially. There are multiple definitions of the PLS algorithm available in the literature, being NIPALS \citep{Wold1973} and SIMPLS \citep{DeJong1993} the most frequently used ones. Here  a version of NIPALS that will be useful in the implementation of the fPQR algorithm is considered:

\begin{enumerate}[Step 1:]
\item Define $X_0=X$ and $Y_0=Y$.
\item Compute $S_a=X_{a-1}^tY_{a-1}$ the sample covariance matrix.
\item Obtain the eigen decomposition of $S_{a}S_{a}^t$ and take $\bm{w}_{a}$ as the eigenvector associated to the largest eigenvalue.
\item Calculate the $X$ score vector as $\bm{t}_{a}=X_{a-1}\bm{w}_{a}$.
\item Calculate the $X$ loading vector as $\bm{p}_{a}=\dfrac{X_{a-1}^t\bm{t}_{a}}{\bm{t}_{a}^t\bm{t}_{a}}$.
\item Calculate the $Y$ loading vector as $\bm{q}_{a}=\dfrac{Y_{a-1}^t\bm{t}_{a}}{\bm{t}_{a}^t\bm{t}_{a}}$.
\item Deflat the matrix $X_{a-1}$ from the information already explained by scores $\bm{t}_1$ and obtain $X_{a}=X_{a-1} - \bm{t}_{a}\bm{p}_{a}^t$.
\item Deflat the matrix $Y_{a-1}$ from the information already explained by scores $\bm{t}_{a}$ and obtain $Y_{a}=Y_{a-1} - \bm{t}_{a}\bm{q}_{a}^t$.

\end{enumerate}

Iterate through steps 2-8 until all $h$ components are computed. Observe that the deflation process stated in step 7 ensures that the score matrix $T$ will be orthogonal. Once all the required components have been computed, the parameter estimates $\hat{\Gamma}$ from equation \eqref{eq:lm_pls} are obtained solving the low dimensional least squares model,
\begin{equation}\label{eq:ols_pls}
\hat\Gamma = \argmin_{\Gamma}\left\{ \| Y-T\Gamma\|^2 \right\}.
\end{equation}

Finally, one can project the estimate $\hat\Gamma$ back into the original sub-space spawned by $X$ and obtain,

\begin{equation}\label{eq:coef_projection}
    \hat{B} = W(P^tW)^{-1}\hat\Gamma.
\end{equation}

PLS is essentially a covariance maximization problem where, at each iteration $a+1$, the objective function being solved is defined as,
\begin{equation}\label{eq:max_cov}
\bm{w}_{a+1} = \argmax _{\bm{w},\|\bm{w}\|=1}\left\{\operatorname{cov}(X_a\bm{w},Y_a) \operatorname{cov}(X_a\bm{w},Y_a)^t\right\},
\end{equation}
where $X_0=X$ and $Y_0=Y$, and the solution is the eigenvector associated to the largest eigenvalue $\lambda_1$,
\begin{equation}
S_aS_a^t\bm{w_a} = \lambda_1\bm{w_a}.
\end{equation}

Posing PLS as a covariance optimization problem opens the door to the possibility of using alternative covariance definitions. Traditionally, robust versions have been considered in order to obtain robustified PLS algorithms, see for example \citep{Hubert2003}. In this work we are interested in defining not only a robust PLS estimator, but an estimator linked to the quantiles of the response matrix, giving the possibility to study the tails of the response matrix and not just the central behavior. As a solution to this question, a robust quantile based  dimension reduction technique that we call fast partial quantile regression (fPQR) is introduced in the next section.


\section{Fast partial quantile regression}\label{sec:fPQR_definition}

There are two key steps in the definition of the fPQR methodology. First, the usage of a quantile covariance metric linked to the quantiles, instead of the traditional covariance, that is linked to the mean. As it will be discussed in Section \ref{sec:numerical_sim}, the metric that we consider to be the best alternative was proposed by \cite{Li2015b}, although other alternatives \citep{Dodge2009}, \citep{Choi2018} will also be studied along Sections \ref{sec:other_qcov} and \ref{sec:numerical_sim}. Second, the estimation of the $\Gamma$ coefficients defined in equation \eqref{eq:lm_pls}. In the PLS algorithm, these coefficients are estimated using ordinary least squares, but in the fPQR algorithm a quantile regression model is used instead ensuring that the $\hat\Gamma$ estimates remain linked to the quantiles of the response matrix $Y$.


\subsection{A quantile covariance}\label{sec:qcov_li2015}

In a very interesting work, \cite{Li2015b} extended the usage of autoregressive models to the quantile framework by defining a novel measure suitable for examining the linear relationships between any two random variables for a  given quantile $\tau\in(0,1)$, a measure that they called quantile correlation. Given two random variables $Z_1$ and $Z_2$, take $\operatorname{Q}_{\tau,Z_2}$ as the $\tau$-th quantile of $Z_2$ and $\operatorname{Q}_{\tau,Z_2}(Z_1)$ as the $\tau$-th quantile of $Z_2$ conditional to $Z_1$. Then it is possible to demonstrate that $\operatorname{Q}_{\tau,Z_2}(Z_1)$ is independent of $Z_1$ if and only if the random variables $I(Z_2-\operatorname{Q}_{\tau,Z_2}>0)$ and $Z_1$ are independent, where $I(\cdot)$ is the indicator function. This fact motivated the definition of the quantile covariance proposed in their work as,
\begin{equation}\label{eq:qcov_li}
\begin{aligned}
\operatorname{qcov}_{\tau}\{Z_1, Z_2\} &=\operatorname{cov}\left\{I\left(Z_2-\operatorname{Q}_{\tau, Z_2}>0\right), Z_1\right\} \\
&=E\left\{\psi_{\tau}\left(Z_2-\operatorname{Q}_{\tau, Z_2}\right)(Z_1-E Z_1)\right\},
\end{aligned}
\end{equation}
where $\psi_{\tau}(w)=\tau-I(w<0)$. Being based on a traditional covariance makes this quantile covariance easy and fast to compute. Additionally, although this definition is proposed for random variables, it can be extended to random vectors, making it possible to adapt to the data matrices found in multidimensional problems. Observe however that, opposed to the traditional covariance, this quantile covariance does not enjoy the symmetry property, that is, $\operatorname{qcov}_\tau(Z_1, Z_2)\neq \operatorname{qcov}_\tau(Z_2, Z_1)$. Nevertheless, the lack of symmetry of the quantile covariance does not affect the algorithm proposed in this work, as it is defined for regression, where the roles of independent and dependent variables are clearly specified and the data matrices do not play a symmetric role. A complete definition of this metric can be found in \citep{Li2015b} where they study a nice relation between this metric and the slope from a quantile regression model, and also the asymptotic properties of the estimator.


\subsection{ The fPQR algorithm}\label{sec:fqr_algorithm}

The objective function that the fPQR algorithm solves is obtained by adapting the objective function from a PLS model as it was defined in equation \eqref{eq:max_cov} using the quantile covariance introduced in Section \ref{sec:qcov_li2015},
\begin{equation}\label{eq:max_qcov}
\begin{aligned}
\bm{w}_{a+1} &= \argmax _{\bm{w},\|\bm{w}\|=1}\left\{\operatorname{qcov}_\tau(X_a\bm{w},Y_a)^t \operatorname{qcov}_\tau(X_a\bm{w},Y_a)\right\} \\
&=\argmax _{\bm{w},\|\bm{w}\|=1}\left\{ \bm{w}^tX_a^t\psi_{\tau}(Y_a-\operatorname{Q}_{\tau,Y_a})\psi_{\tau}(Y_a-\operatorname{Q}_{\tau,Y_a})^tX_a\bm{w}\right\},
\end{aligned}
\end{equation}
where $\psi_{\tau}(w)=\tau-I(w<0)$. The solution to this equation is the eigenvector associated to the largest eigenvalue $\lambda_1$,
\begin{equation}
X_a^t\psi_{\tau}(Y_a-\operatorname{Q}_{\tau,Y_a})\psi_{\tau}(Y_a-\operatorname{Q}_{\tau,Y_a})^tX_a\bm{w_a} = \lambda_1\bm{w_a}.
\end{equation}

Based on this idea, the main steps of the fPQR algorithm are defined below,

\begin{enumerate}[Step 1:]
\item Take $\tau\in(0,1)$ the quantile level of interest.
\item Define $X_0=X$ and $Y_0=Y$.
\item Compute $S_{a,\tau}=\operatorname{qcov}_\tau(X_{a-1},Y_{a-1})$ the sample quantile covariance matrix.
\item Obtain the eigen decomposition of $S_{a,\tau}S_{a,\tau}^t$ and take $\bm{w}_a$ as the eigenvector associated to the largest eigenvalue.
\item Calculate the $X$ score vector as $\bm{t}_{a}=X_{a-1}\bm{w}_{a}$.
\item Calculate the $X$ loading vector as $\bm{p}_{a}=\dfrac{X_{a-1}^t\bm{t}_{a}}{\bm{t}_{a}^t\bm{t}_{a}}$.
\item Calculate the $Y$ loading vector as $\bm{q}_{a}=\dfrac{Y_{a-1}^t\bm{t}_{a}}{\bm{t}_{a}^t\bm{t}_{a}}$.
\item Deflat the matrix $X_{a-1}$ from the information already explained by scores $\bm{t}_1$ and obtain $X_{a}=X_{a-1} - \bm{t}_{a}\bm{p}_{a}^t$.
\item Deflat the matrix $Y_{a-1}$ from the information already explained by scores $\bm{t}_{a}$ and obtain $Y_{a}=Y_{a-1} - \bm{t}_{a}\bm{q}_{a}^t$.
\end{enumerate}

Iterate through steps 2-8 until all $h$ components are computed. In order to obtain the parameter estimates $\hat\Gamma$ in the PLS algorithm, a least squares model was solved following equation \eqref{eq:ols_pls}, but in the fPQR algorithm this is substituted by a quantile regression model solving,
\begin{equation}\label{eq:qr}
\tilde{\Gamma}=\argmin_{\bm{\beta}}\left\{ \frac{1}{n}\sum_{i=1}^n\rho_{\tau}(\bm{y}_i -\bm{t}_i^t\Gamma) \right\},
\end{equation}
where $\rho_{\tau}(u)=u(\tau-I(u<0))$ is the quantile regression loss check function. Using a quantile regression model here ensures that the $\hat\Gamma$ estimates remain linked to the quantile of the response matrix $Y$. Finally, one can project $\hat\Gamma$ back into the original sub-space spawned by $X$ as it was done in the PLS models in equation \eqref{eq:coef_projection}. The fPQR is an algorithm that shares many of the benefits of PLS:
\begin{itemize}
    \item It is a dimension reduction technique suitable for multicollinear or high dimensional data;
    \item The new scores obtained by the algorithm are orthogonal;
    \item It maximizes the quantile covariance between predictor and response.
\end{itemize}
But it also has some additional properties:
\begin{itemize}
\item It is a robust methodology, suitable for dealing with outliers or heteroscedastic data;
\item It can provide an estimation of the central behavior of the response conditional to the predictors, but additionally can provide an estimation of any other quantile of the response, conditional to the predictors, obtaining a complete view of the distribution of the response.
\end{itemize}


\subsection{Other quantile covariance metrics}\label{sec:other_qcov}

In Section \ref{sec:fqr_algorithm}, the fPQR algorithm was defined as an optimization problem where a quantile covariance metric is maximized. Although the metric proposed by \cite{Li2015b} was used in the definition of the algorithm, it is possible to consider alternative versions of fPQR based on other quantile covariance metrics. Along this section, two other candidates, defined by \citep{Dodge2009} and \citep{Choi2018} are considered, showing their definition and some properties related to the fPQR performance. 


\subsubsection{A quantile covariance from \cite{Dodge2009}}\label{sec:qcov_dodge}

Take two random variables $Z_1$ and $Z_2$ following the linear model,
\begin{equation}\label{eq:rv_model}
    Z_2 = Z_1\beta +\varepsilon.
\end{equation}

The analytical solution of the ordinary least squares estimator for model \eqref{eq:rv_model} is,
\begin{equation}\label{eq:cov_ols}
    \hat\beta = \operatorname{var}(Z_1)^{-1}\operatorname{cov}(Z_1, Z_2).
\end{equation}

\cite{Dodge2009} take advantage of this fact and define a quantile covariance in terms of the quantile regression estimator, mimicking the relation between the OLS estimator and the traditional covariance displayed in equation \eqref{eq:cov_ols}. Consider the quantile regression estimator,
\begin{equation}\label{eq:qr_estimator_def}
    \tilde\beta = \argmin_\beta \left\{ \operatorname{E}\rho_\tau(Z_2 - \beta Z_1) \right\},
\end{equation}
where $\rho_{\tau}(u)=u(\tau-I(u<0))$ is the quantile regression loss check function. Then the quantile covariance proposed by \cite{Dodge2009} is obtained as,
\begin{equation}
    \operatorname{qcov}_{\tau}^*(Z_1,  Z_2)  = \operatorname{var}(Z_1)\tilde\beta,
\end{equation}
where $\tilde\beta$ is the quantile regression estimator defined in equation \eqref{eq:qr_estimator_def}. Here the superscript ``$*$'' differentiates this quantile covariance from the one defined in Section \ref{sec:qcov_li2015}. There are some remarks worth mentioning:
\begin{itemize}
    \item The extension of this quantile covariance to a multidimensional setting is not as straightforward as in the traditional covariance or in the quantile covariance proposed by \cite{Li2015b}. Given a random vector $U \equiv (U_1,\ldots,U_m)$, the quantile covariance $\operatorname{qcov}_{\tau}^*$ requires to solve a quantile regression model. If the number of variables in the vector $U$ changes, the value of the quantile regression coefficients associated to the variables would also change, affecting the quantile covariance metric. Additionally, when dealing with a high dimensional scenario this would require solving a high dimensional quantile regression model, which is not feasible. The only way to ensure that the results of this quantile covariance are stable and are also feasible in high dimensional problems is to compute it  univariatedly. This way, the computation of the quantile covariance between $U$ and $Z_2$ requires to solve $m$ univariate quantile regression models, where $m$ is the dimension of $U$, greatly affecting the computation time as the number of variables increase;
    \item As happened with the quantile covariance defined by \cite{Li2015b}, this quantile covariance is not symmetric. This means that $\operatorname{qcov}_{\tau}^*(Z_1, Z_2)\neq \operatorname{qcov}_{\tau}^*(Z_2, Z_1)$
\end{itemize}

Additionally to the quantile covariance described above, the key contribution of \cite{Dodge2009} was the adaptation of the univariate NIPALS algorithm to the quantile regression framework. The main differences between their proposal (PQR) and the work developed here (fPQR) are listed below:
\begin{itemize}
    \item In the work developed here, the optimization problem that the fPQR algorithm solves is clearly defined, and based on this definition, the algorithm is proposed. Opposed to this, \cite{Dodge2009} simply defined the algorithm as a modification of the univariate PLS NIPALS, without studying the optimization problem;
    \item The fPQR algorithm allows $Y$ to be a multivariate response matrix while the PQR algorithm is limited to univariate responses;
    \item As it will be seen in Section \ref{sec:numerical_sim}, the covariance considered in the fPQR algorithm allows the algorithm to run significantly faster than the PQR algorithm.
\end{itemize}


\subsubsection{A quantile covariance from \cite{Choi2018}}

Given two random variables $Z_1$ and $Z_2$, the Pearson correlation between the two variables can be seen as the geometric mean of two OLS slopes, $\beta_{2.1}$ of $Z_1$ on $Z_2$ and $\beta_{1.2}$ of $Z_2$ on $Z_1$,
\begin{equation}
    \operatorname{cor}(Z_1, Z_2) = \operatorname{sign}(\beta_{2.1})\sqrt{\beta_{2.1}\beta_{1.2}}.
\end{equation}

Based on this idea, \cite{Choi2018} proposed a quantile correlation coefficient defined as the geometric mean of two quantile regression slopes,
\begin{equation}\label{eq:qcor_choi}
    \operatorname{qcor}^{**}_\tau(Z_1, Z_2) = \operatorname{sign}(\beta_{2.1}(\tau))\sqrt{\beta_{2.1}(\tau)\beta_{1.2}(\tau)},
\end{equation}
where the superscript ``$**$'' is used to differentiate this metric from the ones from  \citep{Li2015b} and \citep{Dodge2009}. A full review of the properties of this metric can be found in the original paper \citep{Choi2018} but there are some remarks that are worth mentioning:

\begin{itemize}
    \item The computation of this quantile covariance metric relies in solving two quantile regression models. This means that the extension of this metric to a multivariate setting faces the same problems as \citep{Dodge2009}. Given a random vector $U \equiv (U_1,\ldots,U_m)$ in order to ensure the stability and feasibility of the results obtained with this metric, it must be computed univariatedly. The computation of $\operatorname{qcor}_{\tau}^{**}(U, Z_2)$ requires thus to solve $2m$ univariate quantile regression models, where $m$ is the dimension of $U$, greatly affecting the computation time;
    
    \item Opposed to the other quantile metrics under study, this is the only metric that is symmetric, meaning that $\operatorname{qcor}_{\tau}^{**}(Z_1, Z_2) = \operatorname{qcor}_{\tau}^{**}(Z_2, Z_1)$.
\end{itemize}

Take into account that the fPQR algorithm requires a quantile covariance, and not a quantile correlation. Although not defined in the original paper, it is possible to obtain an estimation of a quantile covariance based on equation \eqref{eq:qcor_choi}. Observe that,
\begin{equation}
\begin{aligned}
\operatorname{qcor}^{**}_\tau(Z_1, Z_2)&=\operatorname{sign}\left(\beta_{2,1}(\tau)\right) \sqrt{\beta_{2,1}(\tau) \beta_{1,1}(\tau)}\\
&=\operatorname{sign}\left(\beta_{2,1}(\tau)\right) \sqrt{\frac{\operatorname{qcov}_{\tau}^{*}(Z_1, Z_2) \operatorname{qcov}^{*}_{\tau}(Z_2, Z_1)}{\operatorname{var}(Z_1) \operatorname{var}(Z_2)}},
\end{aligned}
\end{equation}
where $\operatorname{qcov}^{*}(\cdot,\cdot)$ refers to the quantile covariance introduced in Section \ref{sec:qcov_dodge}. This way, a symmetric quantile covariance can be defined as,
\begin{equation}
    \operatorname{qcov}^{**}_\tau(Z_1, Z_2) = \operatorname{sign}(\beta_{2,1}(\tau))\sqrt{\operatorname{qcov}_{\tau}^{*}(Z_1, Z_2) \operatorname{qcov}^{*}_{\tau}(Z_2, Z_1)}.
\end{equation}


\section{Numerical simulation}\label{sec:numerical_sim}

This section shows the performance of the proposed fPQR methodology under different synthetic datasets. The three quantile covariances under study, proposed by \citep{Li2015b}, \citep{Dodge2009} and \citep{Choi2018} are compared here. Additionally, the algorithm is compared against PLS, taken as a benchmark model, and the partial robust adaptive modified maximum likelihood estimator (PRAMML), proposed by \cite{Acitas2020}, which is a robust PLS alternative for univariate response models. In order to compare the quantile estimation provided by fPQR with the mean estimations from PLS and PRAMML, the quantile level of the fPQR is fixed at $\tau=0.5$ (the median estimation). For each dataset $\mathbb{D}$, a partition into two disjoint subsets, $\mathbb{D}_{train}$ and $\mathbb{D}_{test}$ is considered. $\mathbb{D}_{train}$ is used for training the models, this is, solving the model equations. $\mathbb{D}_{test}$ is used for testing the models prediction accuracy. The following metrics are computed, where ``$\#$'' denotes the cardinal of a set:

\begin{itemize}
    \item $\|\hat{\bm{\beta}}-\bm{\beta}\|_2$: the euclidean distance between the estimated coefficients and the true coefficients;
    \item $\dfrac{1}{\#\mathbb{D}_{test}}\sum(\hat{y}_i-y_i)^2$:  the mean squared error between the estimated response and the true response;
    \item $E_{\tau}=\dfrac{1}{\#\mathbb{D}_{test}}\sum\rho_{\tau}(y_i-\bm{x}_i^t\hat{\bm{\beta}})$: a quantile error metric between the estimated response and the true response;
    
    \item The execution time of each algorithm measured in seconds.
\end{itemize}

\noindent\textit{Remark}. These simulations compare the results of the fPQR algorithm with the results from PLS and PRAMML. For this reason, the quantile level is fixed at $\tau=0.5$ and the mean squared error is among the metrics considered. However, when dealing with other quantile levels, the mean squared error is not a suitable metric, as it does not take into account the quantile being computed. In such scenarios the quantile error metric is a better alternative.

\subsection{Simulation 1}\label{sec:sim_alvaro}

The following simulation scheme is an adaptation taken from \cite{Mendez-Civieta2020}. The idea behind this scheme is to simulate the behavior found in the increasingly common problem of sparse high dimensional data, where the number of variables is very large, and not all the variables affect the response, being some of them just noise. This problem can be found in many different areas of scientific research such as genetics \citep{Boulesteix2006} or climate data \citep{Chatterjee2011}, and an interesting solution is the usage of dimension reduction techniques like PLS or the proposed fPQR algorithm. Take the model,
\begin{equation}
    \bm{y} = X\bm{\beta}+\varepsilon,
\end{equation}
where the predictors matrix $X$ is generated from a standard normal distribution and the error term is generated following a chi squared distribution with $3$ degrees of freedom, a distribution known to be heavy tailed and non symmetric. This will favor the usage of robust estimators. Since we are interested in the high dimensional framework, a sample size of $n=100$ training observations and $m=100$ predictive variables is considered. Out of the $100$ predictive variables, $30$ are generated from a standard uniform distribution and the remaining $70$ have value $0$, meaning that these $70$ variables do not affect the response variable and are simply noise in the model. Although in real datasets the number of components in the model should be found based on some sort of cross-validation process, in this simulation it is fixed, taken equal to the number of significant variables, $h=30$. Additionally, a sample of $500$ observations is generated as test set. Observe that this fact does not affect the consideration of the simulation being high dimensional, as the algorithms are trained with a number of observations equal to the number of variables. This data generation process is repeated $100$ times, and the results are summarized in terms of the mean value and standard deviation value (shown in parenthesis) of each metric computed.

\begin{table}
	\centering
	\footnotesize
	\caption{Simulation 1. Sparse high dimensional framework considering a $\chi^2(3)$ error.}
	\begin{tabular}{lllll}
		\toprule
		                 & $\norm{\hat{\bm{\beta}}-\bm{\beta}}$   & $\dfrac{1}{\#\mathbb{D}_{test}}\|\hat{\bm{y}}-\bm{y}\|_2^2$ & $E_{\tau}$ & Execution time \\ \midrule
		fPQR Li          & $3.88$ $(0.58)$   & $21.59$ $(5.13)$   & $1.82$ $(0.21)$ & $0.038$ $(0.01)$  \\
		fPQR Dodge       & $4.05$ $(0.62)$   & $23.02$ $(5.98)$   & $1.88$ $(0.23)$ & $38.65$ $(1.649)$  \\
		fPQR Choi        & $4.95$ $(0.94)$   & $31.48$ $(11.40)$  & $2.20$ $(0.35)$ & $76.78$ $(2.716)$  \\
		PLS              & $8.03$ $(2.03)$   & $75.42$ $(37.21)$  & $3.37$ $(0.79)$ & $0.004$ $(0.001)$  \\
		PRAMML           & $6.64$ $(1.37)$   & $52.11$ $(20.66)$  & $2.82$ $(0.54)$ & $0.358$ $(0.047)$  \\
		\bottomrule
	\end{tabular}
	\label{tab:sim3_results}
\end{table}

Results from this simulation scheme are displayed in Table \ref{tab:sim3_results}. In terms of the euclidean distance of the $\beta$ coefficients, the best results are obtained by the fPQR Li estimator, followed by the other quantile based alternatives, while PLS obtains the worst results, as expected since the normality assumptions are not met. Observe also that the standard deviation of this metric is smallest in the fPQR Li, indicating more stable results. In terms of prediction accuracy both in terms of the mean squared error and the quantile error, the best results are obtained also by the fPQR Li algorithm, followed by the fPQR Dodge and achieving again the smallest standard deviation values. Finally, regarding the execution time the fastest algorithm was PLS and the second fastest was fPQR Li, while PRAMML took on average $10$ times longer than fPQR Li. One can also see the large execution times using fPQR Dodge or fPQR Choi alternatives. This is due to the way these covariances are computed, requiring to solve, at each iteration of the algorithm, $m=100$ univariate quantile regression models in the case of Dodge metric, and $2m=200$ models in the case of Choi metric, as it was discussed in Section \ref{sec:other_qcov}.


\subsection{Simulation 2}\label{sec:sim_multi}
A second simulation is considered where we study the problem of having a multivariate response variable, very common in the field of chemometrics. Take,
\begin{equation}
    Y = XB+\varepsilon,
\end{equation}
where the predictors matrix $X$ of size $n=100$ and $m=100$ is generated from a standard normal distribution, and the matrix of coefficients $B$ has size $m=100$ and $l=3$. This defines a problem where the response matrix $Y$ has $l=3$ dimensions. Out of the $100$ predictive variables, $30$ are generated from a standard uniform distribution and the remaining $70$ have value $0$, and finally the error term is generated following a chi squared distribution with $3$ degrees of freedom. In this simulation, the number of components obtained by the algorithms is taken equal to the number of significant variables, $h=30$. Additionally, a sample of $500$ observations is generated as test set and the simulation is repeated $100$ times to ensure the stability of the results. Algorithms PLS and fPQR can deal directly with multivariate response matrices, but PRAMML solves only univariate models, for this reason in this simulation the predictions from PRAMML are obtained by solving $l=3$ independent univariate models.

\begin{table}
	\centering
	\footnotesize
	\caption{Simulation 2. Sparse high dimensional framework with multidimensional response, considering a $\chi^2(3)$ error.}
	\begin{tabular}{lllll}
		\toprule
		                 & $\norm{\hat{\bm{\beta}}-\bm{\beta}}$   & $\dfrac{1}{\#\mathbb{D}_{test}}\|\hat{\bm{y}}-\bm{y}\|_2^2$ & $E_{\tau}$ & Execution time \\ \midrule
		fPQR Li          & $5.16$ $(0.42)$   & $15.14$ $(1.85)$  & $1.52$ $(0.09)$  & $0.10$ $(0.013)$  \\
		fPQR Dodge       & $6.11$ $(0.48)$   & $16.37$ $(2.18)$  & $1.58$ $(0.10)$  & $116.645$ $(2.139)$  \\
		fPQR Choi        & $6.70$ $(0.51)$   & $21.55$ $(3.89)$  & $1.81$ $(0.16)$  & $232.64$ $(7.088)$  \\
		PLS              & $8.61$ $(1.01)$   & $32.01$ $(6.55)$  & $2.21$ $(0.21)$  & $0.023$ $(0.004)$  \\
		PRAMML           & $12.06$ $(1.38)$  & $56.03$ $(11.74)$ & $2.93$ $(0.30)$  & $1.02$ $(0.063)$  \\
		\bottomrule
	\end{tabular}
	\label{tab:sim_multiple_results}
\end{table}

Results from this simulation scheme are displayed in Table \ref{tab:sim_multiple_results}. The best results in terms of the euclidean distance and prediction error are achieved by the fPQR Li algorithm, closely followed by fPQR Dodge. The fPQR Li algorithm also displays the smallest standard deviations, meaning that the results are stable. The PRAMML estimator is outperformed here by all the other algorithms including PLS, probably due to the inability to directly solve multivariate problems, requiring to solve those in a univariate manner. In terms of execution time, the fastest algorithm is PLS, while fPQR Li is the second fastest running 10 times faster than PRAMML. The fPQR Dodge and Choi algorithms are again the slowest.

\subsection{Simulation 3}\label{sim_paper_2005}

The last simulation considered takes the scheme from \citep{Serneels2005} and \citep{Acitas2020}. Consider the model,
\begin{equation}
    \bm{y} = X\bm{\beta}+\bm{\varepsilon}=TP^{t}\bm{\beta}+\bm{\varepsilon},
\end{equation}
where $X=TP^t\in\mathbb{R}^{n\times m}$ is the predictor matrix, $T\in\mathbb{R}^{n\times h}$ is a scores matrix and $P\in\mathbb{R}^{m\times h}$ is a loadings matrix. $T$ and $P$ are generated based on a $N(0,1)$ distribution, and $\bm{\beta}\in\mathbb{R}^m$ is the vector of true coefficients, generated based on a normal distribution with mean $0$ and standard deviation $0.001$. Three possible error distributions are considered for $\bm{\varepsilon}\in\mathbb{R}^n$: a standard normal distribution, a $t_1$ distribution, which is symmetric as the normal distribution but with heavier tails, and a slash distribution (defined as a standard normal distribution divided by a standard uniform distribution), which is heavy tailed and non symmetric. The number of components in the model is fixed, equal to the dimension of the latent loadings $h$. This process is repeated $500$ times. Two cases are defined based on changes in the number of training observations $n$, variables $m$ and components $h$, 

\begin{itemize}
    \item A low dimensional example: $(n, m, h)=(100, 10, 2)$;
    \item A high dimensional example: $(n, m, h)=(15, 60, 4)$.
\end{itemize}

\begin{table}
	\centering
	\footnotesize
	\caption{Simulation 3. Euclidean distance of $\bm\beta$ coefficient estimations under different error distributions.}
	\begin{tabular}{llll}
		\toprule
		                 & $N(0,1)$              &     $t_1$       & Slash           \\ \midrule
		\multicolumn{4}{c}{$(n,m,h)=(100,10,2)$}                                     \\ \midrule
		fPQR Li          & $0.19$ $(0.13)$    & $0.25$ $(0.15)$  & $0.37$ $(0.23)$   \\
		fPQR Dodge       & $0.19$ $(0.13)$    & $0.26$ $(0.16)$  & $0.38$ $(0.24)$   \\
		fPQR Choi        & $0.49$ $(1.37)$    & $3.46$ $(55.95)$ & $1.69$ $(5.70)$   \\
		PLS              & $0.19$ $(0.10)$    & $6.23$ $(22.57)$ & $12.00$ $(58.51)$ \\
		PRAMML           & $0.16$ $(0.10)$    & $0.23$ $(0.14)$  & $0.31$ $(0.19)$   \\ \midrule
		\multicolumn{4}{c}{$(n,m,h)=(15,60,4)$}                                      \\ \midrule
		fPQR Li          & $0.79$ $(0.33)$    & $1.61$ $(1.25)$    & $2.21$ $(1.45)$ \\
		fPQR Dodge       & $0.90$ $(0.40)$    & $1.84$ $(1.56)$    & $2.49$ $(1.70)$  \\
		fPQR Choi        & $6.74$ $(42.19)$   & $18.78$ $(176.08)$ & $28.25$ $(231.58)$  \\
		PLS              & $1.14$ $(0.42)$    & $14.94$ $(68.17)$  & $29.55$ $(183.84)$  \\
		PRAMML           & $0.61$ $(0.31)$    & $1.02$ $(0.62)$    & $1.42$ $(0.98)$  \\ 
		\bottomrule
	\end{tabular}
	\label{tab:sim_acitas_mse_beta}
\end{table}

\begin{table}
	\centering
	\footnotesize
	\caption{Simulation 3. Execution time}
	\begin{tabular}{lllll}
		\toprule
		fPQR Li   & fPQR Dodge & fPQR Choi & PLS     & PRAMML   \\ \midrule
		\multicolumn{5}{c}{$(n,m,h)=(100,10,2)$}                                     \\ \midrule
		$0.015$   & $0.27$ & $0.54$ & $0.0006$ & $0.017$ \\ \midrule
		\multicolumn{5}{c}{$(n,m,h)=(15,60,4)$}                                       \\ \midrule
		$0.017$ & $2.99$ & $5.94$ & $0.0007$ & $0.021$ \\
		\bottomrule
	\end{tabular}
	\label{tab:sim_acitas_exec_time}
\end{table}

Results from this simulation are shown in Tables \ref{tab:sim_acitas_mse_beta} and \ref{tab:sim_acitas_exec_time}. In terms of the euclidean distance of the $\beta$ coefficients, one can see that PRAMML estimator obtains the best results closely followed by fPQR Li and Dodge algorithms, being both competitive alternatives. It is worth remarking the fact that fPQR Li and Dodge outperformed PLS even when considering a normal distribution for the error term, where PLS is expected to excel. Finally, fPQR Choi consistently provides the worst results. The execution time is affected by the number of observations $n$, variables $m$ and $l$, and components $h$, but not by the error distribution, for this reason Table  \ref{tab:sim_acitas_exec_time} shows the execution time regardless of the error distribution. PLS is the fastest algorithm, while fPQR Li is the second fastest closely followed by PRAMML. Results regarding prediction accuracy are not included in this simulation scheme because the error distributions considered generated outliers with very large values, providing predictions where the mean squared error values were very large and very similar regardless of the algorithm.

The three simulations displayed in this section remark the fact that, among the three quantile covariances under study, the best alternative for the fPQR algorithm is the quantile covariance proposed by \cite{Li2015b}, as it consistently provides the smallest prediction errors and the smallest euclidean distance of the $\beta$ coefficients. Additionally, it is by far the fastest of the three algorithms, having a computation based on a traditional covariance rather than in solving univariate quantile regression models, as is the case with the other quantile covariances considered. Comparing the fPQR Li algorithm for the median with PLS shows that it outperformed PLS in all the scenarios considered in terms of prediction accuracy and euclidean distance of the $\beta$ coefficients. When comparing it with robust PLS alternatives like PRAMML, it is worth remarking the fact that fPQR Li can be used to solve multidimensional response problems while PRAMML requires to face this situation by solving univariate models, as discussed in Section \ref{sec:sim_multi}. Additionally, one can see that fPQR Li is a competitive alternative in terms of prediction accuracy and euclidean distance of the $\beta$ coefficients, providing better estimations in two of the three simulations, and being competitive in the last one. In terms of execution time, fPQR Li also outperformed PRAMML in all the simulations. But the fPQR algorithm has an additional advantage when compared with any PLS based methodology: PLS based methodologies can only obtain estimations for the mean of the response matrix, while fPQR can obtain estimations for different quantile levels. This allows to study not only the central behavior of the response variable, but also the behavior at any other quantile of interest, like the tails of the distribution.


\section{Real data analysis: Biscuit data}\label{sec:real_data}

\begin{figure}
	\centering
	\caption{Biscuit dataset: NIR spectra of the biscuit dataset.}
	\includegraphics[width=12cm]{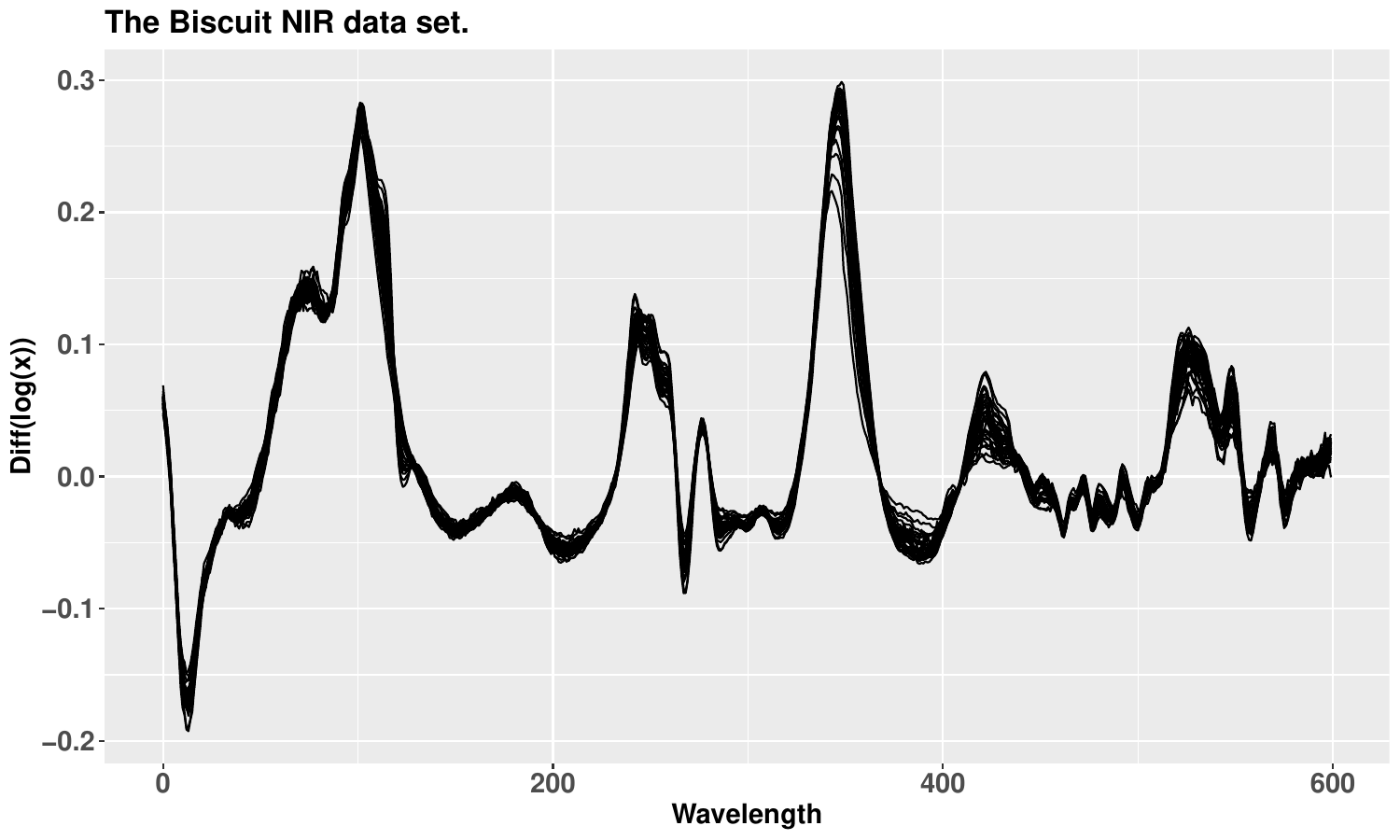}
	\label{fig:nir_spectra}
\end{figure}

\begin{figure}
	\centering
	\caption{Biscuit dataset: Response matrix distribution.}
	\includegraphics[width=12cm]{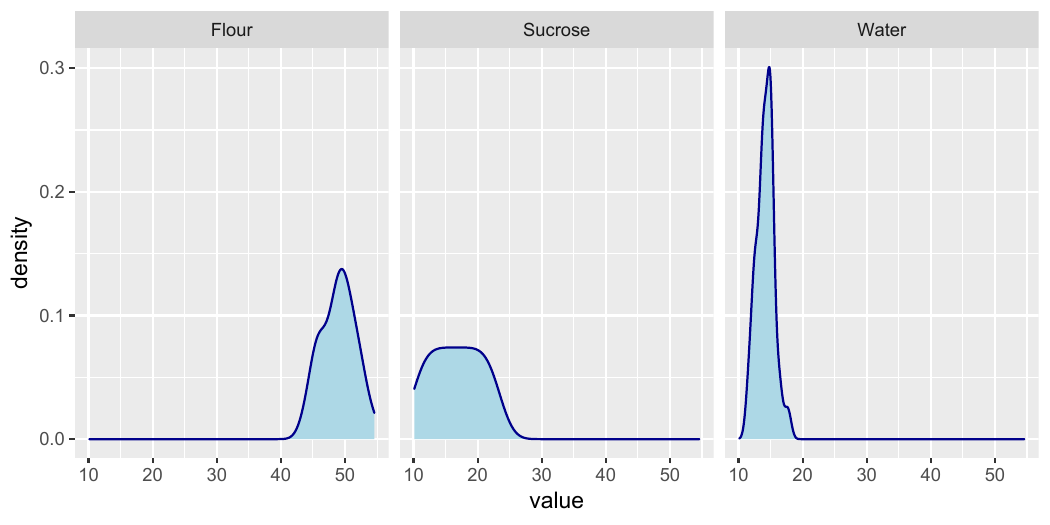}
	\label{fig:response_distribution}
\end{figure}

The biscuit data was first introduced in \cite{Osborne1984}. This dataset contains four response variables, concentration of fat, flour, sucrose and water, of $72$ biscuit dough samples, where $40$ observations usually define a training set and $32$ a prediction set. In this analysis, and following the steps from \citep{Hubert2003}, the variable fat was removed because it showed small correlation coefficients with the other constituents and a larger variance. The rest of the response variables show larger correlations and similar variances, and for this, a multivariate analysis is considered. The objective is to predict the values of the three response variables based on NIR spectra measurements taken every $2$ nm from 1200 up to 2400. The same preprocessing steps as in \citep{Hubert2002} and \citep{Hubert2003} were performed, obtaining a NIR spectra prediction matrix of $m=600$ dimensions, shown in Figure \ref{fig:nir_spectra}, and a response matrix of $l=3$ dimensions whose distribution is shown in Figure \ref{fig:response_distribution}. Observe that the response variables are non normal. Actually, the three of them show certain degree of skewness and are heavy tailed, scenario where algorithms based on normality assumptions like PLS may face problems. Observation $23$ is known to be an outlier, \citep{Osborne1984} suggests that the compositional data of this observation is in error. However, following the steps from \citep{Hubert2003}, it is kept in the dataset.

\begin{figure}
	\centering
	\caption{Biscuit dataset: CV mean squared error on the number of components.}
	\includegraphics[width=12cm]{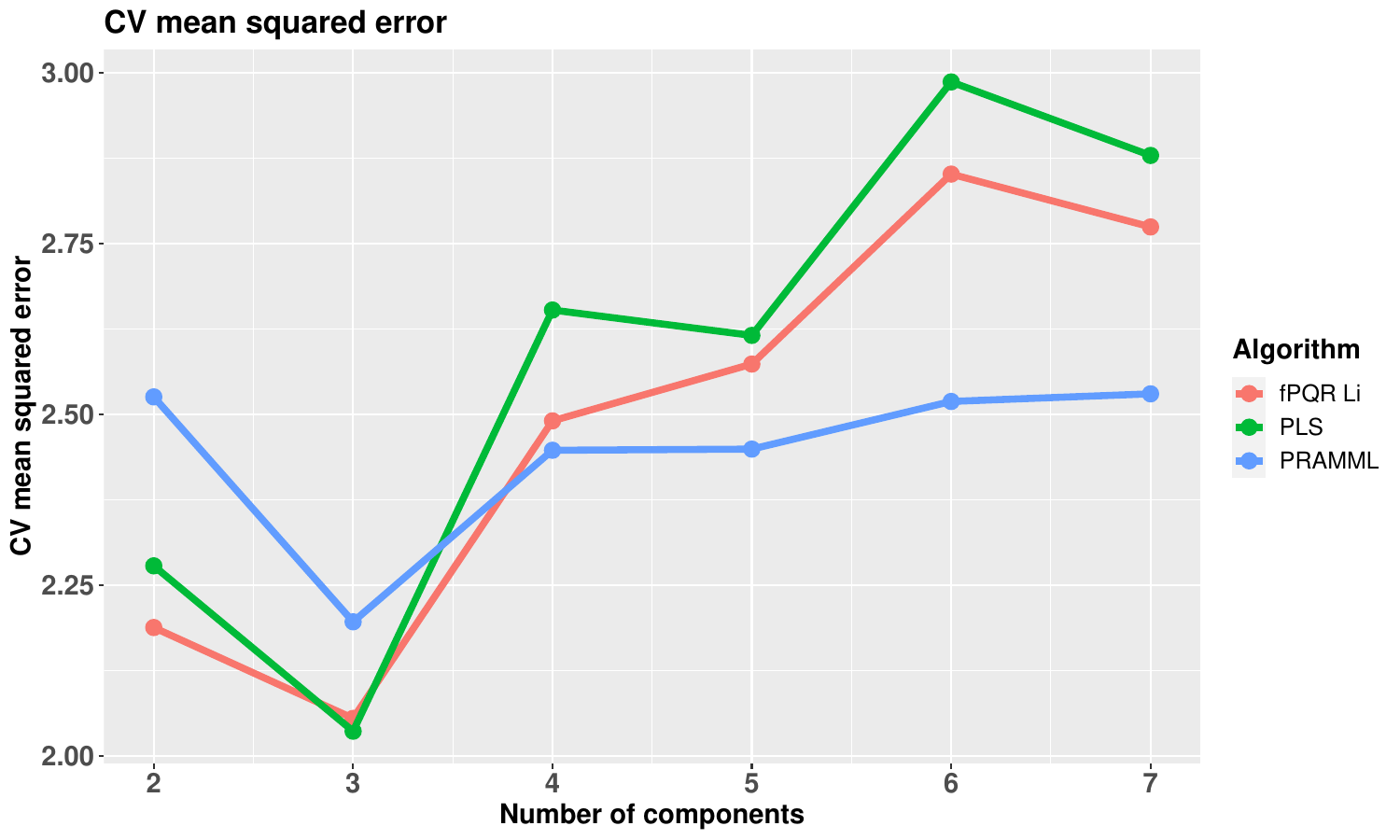}
	\label{fig:cv_biscuit}
\end{figure}

Using this dataset, a comparison of fPQR Li, PLS and PRAMML estimators is performed. The quantile level is taken as $\tau=0.5$ so that quantile based results can be compared with the mean based results from PRAMML and PLS. Since the PRAMML estimator solves only univariate models, the predictions from this estimator are obtained by solving $3$ independent univariate models. The first step is to select the number of components to be computed. This is done by performing $5$-fold cross validation on the training set, and the objective is to minimize the mean squared error of the predictions. A range of possible number of components going from $2$ up to $7$ components is considered here. Although it is more common to start the range of the number of components at $1$, here the starting number of components was limited to $2$ by the PRAMML algorithm, as this algorithm requires a minimum number of $2$ components in order to be executed. Figure \ref{fig:cv_biscuit} shows the CV results, concluding that three is the best number of components for any of the models considered.

\begin{table}
	\centering
	\footnotesize
	\caption{Biscuit data: Test mean squared and quantile regression errors.}
	\begin{tabular}{lll}
		\toprule
		                 & $\dfrac{1}{\#\mathbb{D}_{test}}\|\hat{\bm{y}}-\bm{y}\|_2^2$ & $E_{\tau}$   \\ \midrule
		\multicolumn{3}{c}{Outlier in training set}                                     \\ \midrule
		fPQR Li & $0.491$ & $0.25$  \\
		PLS     & $0.614$ & $0.29$  \\
		PRAMML  & $0.527$ & $0.25$ \\\midrule
		\multicolumn{3}{c}{Outlier in test set}  \\ \midrule
		fPQR Li & $1.93$ & $0.33$ \\
		PLS     & $1.87$ & $0.32$ \\
		PRAMML  & $2.11$ & $0.34$ \\
		\bottomrule
	\end{tabular}
	\label{tab:biscuit_final_ms_result}
\end{table}

As was mentioned above, observation $23$ is known to be an outlier. However, typically one does not know if there are outliers in the dataset or, if any, if those are in the training set or the test set. For this reason here two final models are built. The first includes observation $23$ in the training set and addresses the more complicated problem of having an outlier in the training part. If a model has good predictive results under this circumstance, it implies that it is capable of generalizing correctly, not being influenced by the outlier while making predictions. The second model includes observation $23$ in the test set. Both models are built using the optimal number of components found before, three, and the mean squared error and quantile error of the prediction of each model are computed on the test set. Table \ref{tab:biscuit_final_ms_result} shows the results. When the outlier is included in the training set, one can see that the best result is obtained by fPQR Li, followed by the PRAMML estimator, and PLS obtains the worst result. This suggests that fPQR is the most robust alternative, not influenced by the presence of the outlier while building the model. If the outlier is in the test set, the best results are obtained by PLS, closely followed by fPQR, while PRAMML obtains the worst results. This suggests that fPQR is a good compromise for dealing with outliers in any of the two scenarios.

An additional advantage of fPQR Li is that it can provide estimations for different quantile levels. Take for example the first model and observation $41$, which is the first one in the test set. This observation has values flour$=16.44$, sucrose$=47.65$ and water$=12.57$, and the median prediction obtained using fPQR Li for $\tau=0.5$ is flour$=15.68$, sucrose$=48.39$ and water$=12.82$. But one can also calculate an estimation of any other quantile of interest, obtaining this way prediction intervals. For example, the prediction for the $10\%$ percentile of the response is flour$=15.24$, sucrose$=47.67$ and water$=12.39$ for a small biscuit dough given the associated NIR spectra values, while the $90\%$ percentile for a large biscuit dough has values flour$=17.22$, sucrose$=48.41$ and water$=13.10$. The fPQR Li algorithm can thus provide a complete picture of the distribution of the response matrix.


\section{Computational aspect}\label{sec:comput_aspect}

All the simulations and analysis commented in Sections \ref{sec:numerical_sim} and \ref{sec:real_data} were run in a computer with an Intel Core i7-10750H CPU (2.6GHz) processor with 32GB RAM memory running the O.S. Windows $10$. The computation of the fPQR has been developed in Python 3.8.5 (Anaconda Inc.). The quantile covariance metrics introduced in Section \ref{sec:other_qcov} required solving quantile regression models. Those were solved using the Python package ASGL, built on top of the CVXPY optimization framework for Python \citep{Diamond2016} and Mosek solver \citep{Mosek2021}. The PRAMML estimator was computed using the R package `rpls' \citep{Acitas2020b}, as there was no Python implementation for this methodology.


\section{Conclusion}\label{sec:conclusion}

In this paper the fast partial quantile regression (fPQR) algorithm has been introduced. This algorithm extends the PLS models to the quantile regression framework. The result is a dimensionality reduction technique that parallels the nice properties of PLS models but that is linked to the quantiles of the response matrix, being robust to the presence of outliers and heteroscedastic data. As discussed in Section \ref{sec:fPQR_definition}, the key idea behind fPQR is the definition of the objective function that it maximizes in terms of a quantile covariance metric, and in this work different metrics are considered \citep{Li2015b}, \citep{Dodge2009}, \citep{Choi2018}. Section \ref{sec:numerical_sim} studies the performance of the fPQR algorithm using the different quantile metrics in a set of synthetic datasets, concluding that the best results in terms of prediction accuracy, euclidean distance of the $\beta$ coefficients and execution time are obtained using the quantile covariance defined by \cite{Li2015b}. Additionally, the performance of the fPQR algorithm is compared with PLS and PRAMML \citep{Acitas2020} estimators, showing that, if the median estimation is computed, fPQR is a competitive alternative to other robust PLS algorithms, but additionally, fPQR can obtain estimates for different quantile levels of the response matrix, providing a complete picture of its distribution. The performance of the proposed work is also studied in a real high dimensional dataset containing NIR spectra measurements, where fPQR Li obtains the best results.


\section{Acknowledgments}

This research was partially supported by research grants and projects PID2020-113961GB-I00 and PID2019-104901RB-I00 from Agencia Estatal de Investigación.


\bibliography{references}

\end{document}